\documentclass[aps,prl,twocolumn,showpacs]{revtex4}

\usepackage{graphicx}
\usepackage{dcolumn}
\usepackage{amsmath}

\begin{document}

\title{Magnetic field induced dehybridization of the electromagnons in multiferroic TbMnO$_3$}

\author{P. Rovillain$^1$}
\author{M. Cazayous$^1$}
\author{Y. Gallais$^1$}
\author{M-A. Measson$^1$}
\author{A. Sacuto$^1$}
\author{H. Sakata$^2$}
\author{M. Mochizuki$^3$}

\affiliation{$^1$Laboratoire Mat\'eriaux et Ph\'enom\`enes Quantiques (UMR 7162 CNRS), Universit\'e Paris Diderot-Paris 7, 75205 Paris cedex 13, France\\
$^2$Department of Physics, Tokyo University of Science, 1-3 Kagurazaka Shinjyuku-ku Tokyo 162-8601, Japan\\
$^3$Department of Applied Physics, The University of Tokyo, 7-3-1, Hongo, Bunkyo-ku, Tokyo 113-8656, Japan}

     
\begin{abstract}
We have studied the impact of the magnetic field on the electromagnon excitations in TbMnO$_3$ crystal. 
Applying magnetic field along the c axis, we show that the electromagnons transform into pure antiferromagnetic modes, losing their polar character. 
Entering in the paraelectric phase, we are able to track the spectral weight transfer from the electromagnons to the magnon excitations and we discuss the magnetic excitations underlying the electromagnons. 
We also point out the phonons involved in the phase transition process. This reveals that the Mn-O distance plays a key role in understanding the ferroelectricity and the polar character of the electromagnons. Magnetic field measurements along the b axis allow us to detect a new electromagnon resonance in agreement with a Heisenberg model.   
\end{abstract}

\maketitle

Ferroelectric order is unlikely to be present in magnetic compounds. 
The coexistence of magnetism and ferroelectricity is so rare that the rediscovery of materials combining both properties in the so-called multiferroic materials has given rise to a strong revival of this field. In multiferroics, the coupling between the magnetic and ferroelectric orders represents a promising route to the electrical control of magnetic properties for storage and for processing devices[\onlinecite{Eerenstein, Chu, Baek}]. 
  Some multiferroics not only present a simple phase coexistence between both orders but a ferroelectricity induced by the magnetic structure. 
This is the case for perovskite manganites RMnO$_3$ (R=Rare earth) in which the spiral spin structure breaks the inversion symmetry of the magnetic ordering and gives rise to a ferroelectric state. A purely electronic mechanism or a magnetically induced ionic displacement mechanism based on the Dzyaloshinskii-Moriya coupling have both been proposed for the origin of this ferroelectricity [\onlinecite{Katsura, Sergienko}]. Such mechanisms explain the static properties of perovskites but fail in the interpretation of the dynamical coupling between spin and charge degrees of freedom. 
The mixing of spin waves (magnons) with optical phonons creates a novel class of excitations called electromagnons, magnons with a dipolar activity. 
These electromagnon excitations were recently interpreted using microscopic approaches based on Heisenberg interaction [\onlinecite{Valdes}] and taking into account the ground state spin spiral configuration [\onlinecite{Sousa, Mochizuki}]. 

TbMnO$_3$ is the most representative compound of this class of multiferroics. In its antiferromagnetic phase (T$_N$ = 41 K), a sinusoidal spin structure is replaced by a cycloid magnetic order below 28 K breaking the inversion symmetry of the crystal and inducing a spontaneous polarization along the \textit{c} axis. In this phase, the spins form a spiral with an incommensurate wavevector \textbf{Q$_b$}=(0,0.28,0) in the \textit{bc} plane. The spin-spiral plane can flip from the \textit{bc} into the \textit{ab} plane with an external magnetic field along the \textit{a} and \textit{b} axis. Two electromagnons have been measured  by infrared (IR) [\onlinecite{Sushkov, Takahashi}], terahertz (THz) [\onlinecite{Pimenov}] and Raman [\onlinecite{Rovillain}] spectroscopies and by inelastic neutron scattering [\onlinecite{Senff}]. Both electromagnons are always optically excited using an electric field of the light parallel to the \textit{a} axis irrespective of the spiral-plane orientation (\textit{bc} or \textit{ab}). 
This specific selection rule shows that electromagnons are tied to the lattice rather than the spin structure [\onlinecite{Valdes}]. 
However, the structural origin of the polar activity of the electromagnons has not been yet identified. 

In this letter, we show that the electric dipole character of the electromagnons disappears and their pure magnetic part emerges when the spin spiral is destabilized with a magnetic field along the \textit{c} axis. The spectral weight transfer from the electromagnons to the magnon excitations sheds light on the magnetic excitations involved in the formation of electromagnons. 
The effect of the phase transitions on the phonon modes shows that the Mn-O distance is the crucial length for understanding the ferroelectricity and the polar character of the electromagnons. The phase diagram for a magnetic field along the \textit{b} axis is also explored and a new electromagnon resonance is observed in agreement with calculations from a Heisenberg model. 

\par
Single \textit{Pbnm} crystals of TbMnO$_3$ were grown by the floating-zone method and oriented by Laue x-ray diffraction. Two samples were investigated :  (i) an \textit{ab} and (ii) an \textit{ac} sample for magnetic field along the \textit{c} and \textit{b} axis, respectively. 
We have performed Raman measurements in a backscattering geometry with a triple spectrometer Jobin Yvon T64000 using the 568~nm excitation line from a Ar$^+$-Kr$^+$ mixed gas laser [\onlinecite{Rovillain}]. 

\par
\begin{figure}
\includegraphics*[width=8cm]{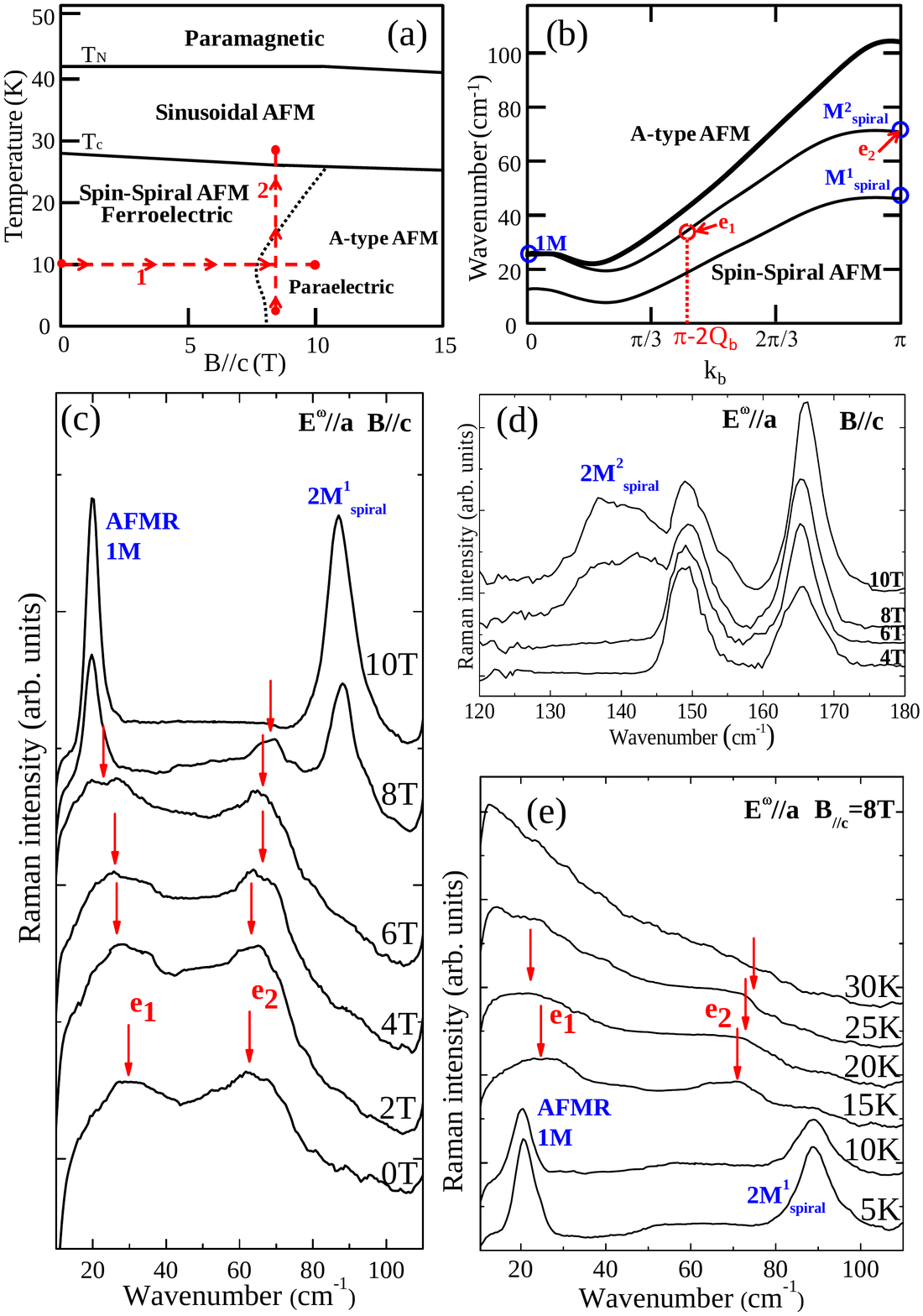}\\
\caption{\label{Figure1} 	 
(a) Magnetoelectric phase diagram of TbMnO$_3$ with \textbf{B}//\textit{c}. 
(b) Magnon dispersions along \textbf{k=}(0, k$_b$,0). The two lower magnon dispersions in the spin-spiral AFM state are extracted from neutron measurements [\onlinecite{Senff2}]. 
The upper (bold) magnon dispersions have been calculated in the A-type AFM state.  The center
1M, M$^1_{spiral}$ and M$^2_{spiral}$ correspond to the zone center magnon and to the zone-edge magnon modes in the spin-spiral state, respectively. 
e$_1$ is the predicted electromagnon with a wavevector equal to $\pi$-2Q$_b$ and e$_2$ the electromagnon at the zone edge of the Brillouin zone.
(c-d) Raman spectra obtained with E$^\omega$//a and B//c along the path 1 (see (a)). 
(e) Raman spectra obtained increasing temperature along the path 2 (see (a)). Arrows indicate the e$_1$ and e$_2$ modes. One can notice that the Tb crystal field at 4.5~meV (36~cm$^{-1}$) is not observed.[\onlinecite{Senff}]}
\end{figure}

Figure~\ref{Figure1}a shows the magnetoelectric phase diagram of TbMnO$_3$ for \textbf{B}//\textit{c} [\onlinecite{Kimura, Argyriou, Mochizuki2}]. 
Previous transport measurements have shown that the spontaneous polarization is rapidly suppressed above 8 T and TbMnO$_3$ becomes paraelectric [\onlinecite{Kimura}]. Additional neutron measurements show that the incommensurate magnetic structure is destabilized between 7 and 10 T [\onlinecite{Argyriou}]. Above 10 T, TbMnO$_3$ becomes a simple A-type antiferromagnetic [\onlinecite{Argyriou}].

Figure~\ref{Figure1}c shows at zero magnetic field the two electromagnons modes (e$_1$$\approx$ 30~cm$^{-1}$ and e$_2$$\approx$ 60~cm$^{-1}$) measured on the \textit{ab} TbMnO$_3$ sample. The electric field of light \textbf{E$^{\omega}$} of the incident and scattered photons are parallel to the \textit{a} axis. These excitations disappear above the Curie temperature indicating their electromagnon origin [\onlinecite{Rovillain}]. In addition THz measurements have detected electromagnons in the same energy ranges and with the same polarization \textbf{E$^{\omega}$}//a further reinforcing this assignment [\onlinecite{Pim}].
The e$_1$ electromagnon is observed in IR and THz at lower energy around 20-25~cm$^{-1}$ which is identical to the energy of the antiferromagnetic resonance (AFMR). These excitations can be distinguished only by their optical selection rules [\onlinecite{Sushkov}]. IR and THz measurements lead to the interpretation of the one magnon (AFMR) labeled 1M in Fig.~\ref{Figure1}b as the magnetic part of the e$_1$ electromagnon. However, recent theoretical approaches predict different energies for the AFMR and the e$_1$ electromagnon (see the dispersion curve in the spiral spin state of Fig.~\ref{Figure1}b) [\onlinecite{Sousa, Mochizuki}]. 
Based on calculations of Refs.~\onlinecite{Sousa} and \onlinecite{Mochizuki}, the wavevectors of the e$_1$ and e$_2 $ electromagnons should be equal to q$_{e_1}$=$\pi$-2Q$_b$ (Q$_b$=0.28 is the magnitude of the cycloidal wavevector) and q$_{e_2}$=$\pi$ (zone-edge electromagnon)
in the dispersion of the spin spiral AFM state (see Fig.~\ref{Figure1}b). 
This dispersion corresponds to the spin oscillations out of the \textit{bc} spiral  plane rotating around the c axis [\onlinecite{Senff}].
Given these discrepancies, it is fundamental to experimentally determine the magnetic excitation associated to the e$_1$ electromagnon in order to find the mechanism at the origin of the electromagnons. 
The investigation of the TbMnO$_3$ phase diagram when a magnetic field \textbf{B} is applied along the \textit{c} axis gives the opportunity to shed light on this question.

Figure~\ref{Figure1}c and d present the Raman spectra  along path 1 (T=10K in Fig.~\ref{Figure1}a). 
The frequency of the e$_1$ and e$_2$ electromagnons (Fig.~\ref{Figure1}c) respectively decreases and increases with the magnetic field.    
For B=8 T, one sharp peak appears at 21~cm$^{-1}$ (1M) in Fig.~\ref{Figure1}c and two broader peaks at 88~cm$^{-1}$ (2M$^1_{spiral}$,Fig.~\ref{Figure1}c)  and 140~cm$^{-1}$ (2M$^2_{spiral}$) in Fig.~\ref{Figure1}d. At the same time, the electromagnons intensity is drastically reduced and disappears at B=10 T. 

The additional peaks measured above 8 T in the paraelectric phase are associated to magnetic excitations. Simple Tb$^{3+}$ f-f transitions have different energies and can be ruled out [\onlinecite{Konig}]. 
The 1M peak corresponds to the zone center magnon of the magnon dispersion in the spin spiral state (see Fig.~\ref{Figure1}b).
The 2M$^1_{spiral}$ and 2M$^2_{spiral}$  peaks are associated to two-magnon excitations. The two-magnon excitation is formed from two zone-edge magnons
 with opposite wavevectors.
The energies of 2M$^1_{spiral}$ and 2M$^2_{spiral}$ two-magnons are twice the M$^1_{spiral}$ and M$^2_{spiral}$ zone edge energies of the magnon dispersions in the spin spiral state (see Fig.~\ref{Figure1}b). The energies of these modes are in agreement with neutron measurements [\onlinecite{Senff2}].

From these observations, one can deduce that the spin structure associated with the spiral spin state survives above 8 T. A magnetic field above 10 T should swing TbMnO$_3$ to a pure A-type antiferromagnetic phase [\onlinecite{Argyriou}] and  
the two-magnon excitation of the zone-edge A-type AFM dispersion should be measured around 200~cm$^{-1}$ (see Fig.~\ref{Figure1}b). This peak is not experimentally detected up to 10 T. The A-type AFM phase was not observed here. The 8-10 T range might be interpreted as a hysteresis region of the first-order phase transition.  
Entering in the paraelectric phase, the electromagnons e$_1$ and e$_2$ disappear quickly whereas the intensity of the one-magnon (1M) and two-magnon excitations (2M$^1_{spiral}$ and 2M$^2_{spiral}$ peaks) drastically increase. 
There is a clear difference between the magnon modes and the electromagnons. Both excitations do not have the same energy and nor the same width. Moreover, there is a clear spectral weight transfer from electromagnons to pure magnon excitations entering the paraelectric phase (see Fig.~\ref{Figure1}c).  
We can notice that the energy of an excitation at $\pi$-2Q$_b$ in the spin spiral AFM dispersion of Fig.~\ref{Figure1}b is around 30~cm$^{-1}$ and the one magnon excitation has an energy around 20~cm$^{-1}$ as experimentally observed in Raman measurements.  
This difference suggests that the e$_1$ electromagnon observed using Raman scattering is associated with a magnetic excitation at $k$=$\pi$-2Q$_b$ and not at $k$=0.  

According to Refs. \onlinecite{Sousa} and \onlinecite{Mochizuki} the $e_2$ electromagnon is Raman active due to the folding of the Brillouin zone. 
The periodicity of the lattice is doubled and the originally $q_{e_2}$=(0,$\pi$,0) electromagnon becomes an excitation at the Gamma ($k$=0) point. 
For the lower energy electromagnon at $q_{e_1}$=(0,$\pi$-2Q$_b$,0), the Brillouin zone is folded at the wave vector equal to Q$_b$ due to the spiral magnetic structure. Then the $e_1$ electromagnon also becomes an excitation at the Gamma point.

Figure~\ref{Figure1}e shows the Raman spectra obtained by increasing the temperature along path 2 (B=8 T) in Fig.~\ref{Figure1}a. The AFMR and the 2M$^1_{spiral}$ peaks disappear between 10K and 15K (the 2M$^2_{spiral}$ is not shown here but it desappears in the same temperature range). 
In the same range of temperatures, the $e_1$ and $e_2$ electromagnon peaks are emerging and then disappear above 25 K. These observations are in agreement with the reported phase diagram (Fig.~\ref{Figure1}a). It confirms that the electromagnons exist only in the ferroelectric phase and that the pure magnetic excitations exist only in the paraelectric phase.   

From the structural point of view, IR measurements have suggested that the Mn atomic displacement is mainly at the origin of the ferroelectricity induced by the Dzyaloshinskii-Moriya interaction [\onlinecite{Lobo}]. For the polar part of the electromagnons, the spectral weight transfer observed in IR from phonons to the electromagnons shows that their polar activity mainly comes from phonons [\onlinecite{Takahashi, Pimenov2, Lobo}]. However the exact structural origin of the ferroelectricity and the origin of the electromagnon polar activity are not determined. 

\begin{figure}
\includegraphics*[width=7cm]{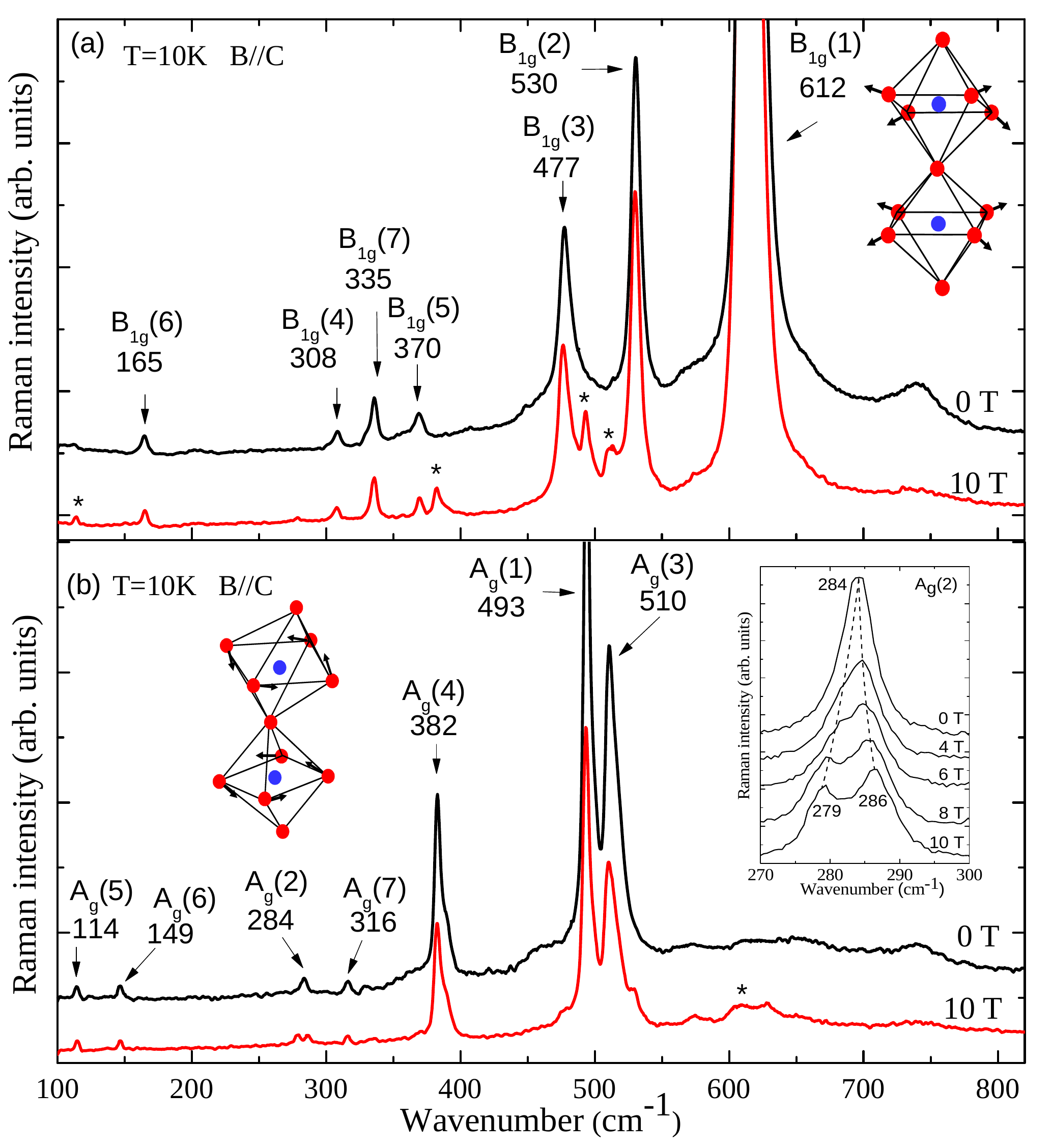}\\
\caption{\label{Figure2} 
Raman spectra of phonon modes for 0 and 10 T in cross $xz$ (a) and parallel $xx$ (b) polarization of light. 
Raman spectra in cross polarization shows 7 B$_{1g}$ modes (a) and 7 A$_g$ modes (b) in parallel polarization at 0 T. The insert in (b) show the splitting of A$_g$(2) phonon mode (MnO$_6$ block rotation) with the magnetic field. The A$_g$  modes are obtained in $z(xx)\overline{z}$ geometry with eigenvectors parallel to the x direction. The B$_{1g}$ modes are obtained in $z(xy)\overline{z}$ geometry with eigenvectors parallel to the x and y directions. 
Stars indicate phonon modes due to polarization leakage.}
\end{figure}

Figure~\ref{Figure2}a and b show the phonon spectra for zero and 10 T along the $c$ axis using cross \textit{xy} (B$_{1g}$) and parallel \textit{xx} (A$_g$) polarizations of light, respectively. Over the 7 A$_g$ and 7 B$_{1g}$  Raman active modes predicted by the group theory, all the modes have been detected and assigned in agreement with previous measurements [\onlinecite{Laverdiere, Iliev}]. The new modes have been assigned by comparison with simulations [\onlinecite{Choithrani}]. A magnetic field along the \textit{a} and \textit{b} axis induces a linear shift of phonon frequencies of less than 0.5~cm$^{-1}$. With a magnetic field along \textit{c}, two phonon modes depart from this simple behavior: the frequency of B$_{1g}$(1) mode shifts of 2~cm$^{-1}$ and the A$_g$(2) mode splits into two peaks (inset of Fig.~\ref{Figure2}b). 

The A$_g$(2) phonon mode is associated to the rotation of the MnO$_6$ block around the \textit{y} axis (cubic notations) (see  Fig.~\ref{Figure2}b).
The Ag(2) phonon mode frequency is linearly proportional to the MnO$_6$ block rotation angle by a factor of 23.5~cm$^{-1}$/deg [\onlinecite{Iliev}].
The splitting of the A$_g$(2) peak (284~cm$^{-1}$) into the 279 and 286~cm$^{-1}$ peaks is equivalent to the rotation of the MnO$_6$ block by -0.2$^o $ and +0.08$^o$, respectively. Moreover, the MnO$_6$ block rotation $\alpha$ is connected to variation of the oxygen atom position: $\alpha=arctan(2\left|(x_0-z_0)\right|)$. The measured rotation of the MnO$_6$ block can be associated with the variation of the oxygen atom position $x_0-z_0$ by about 1.5\% from its initial position.   
We note that the symmetry of the paraelectric phase is higher than the ferroelectric one and less Raman modes are expected in the paraelectric phase. However, TbMnO$_3$ is not a standard ferroelectric and this unusual behaviour has been already pointed out [\onlinecite{Lobo}]. 

The B$_{1g}$(1) mode corresponds to the in-plane oxygen stretching vibration in the xz plane and is determined by the Mn-O distance. This mode is the so called Jahn-Teller mode. The frequency of the B$_{1g}$(1) mode is proportional to $d^{3/2}_{Mn-O}$. The frequency shift of the B$_{1g}$(1) peak corresponds to an increase of 0.1\% of the Mn-O distance (the Mn-O distance is equal to 2.063 $\dot{A}$ in TbMnO$_3$ at zero field). 
Pure electronic or magnetically induced ionic displacement mechanisms have been proposed for the origin of the ferroelectric polarization. 
From an experimental point of view, no phonon anomaly has been detected by inelastic neutron scattering supporting the electronic origin of the ferroelectricity [\onlinecite{Kajimoto}]. Previously, only IR measurements have been able to detect a small phonon softening (0.5~cm$^{-1}$) at the ferroelectric transition [\onlinecite{Lobo}]. Only with the magnetic field along the c-axis, a strong renormalization of two phonons frequency can be measured when TbMnO$_3$ becomes paraelectric and the electric activity of the electromagnons vanishes with a 8 T magnetic field. Based on these observations, we suggest that the ferroelectricity and the polar activity of the electromagnon are related to the change of the oxygen atom position and correlated with the increase of the Mn-O distance. 

\begin{figure}
\includegraphics*[width=8cm]{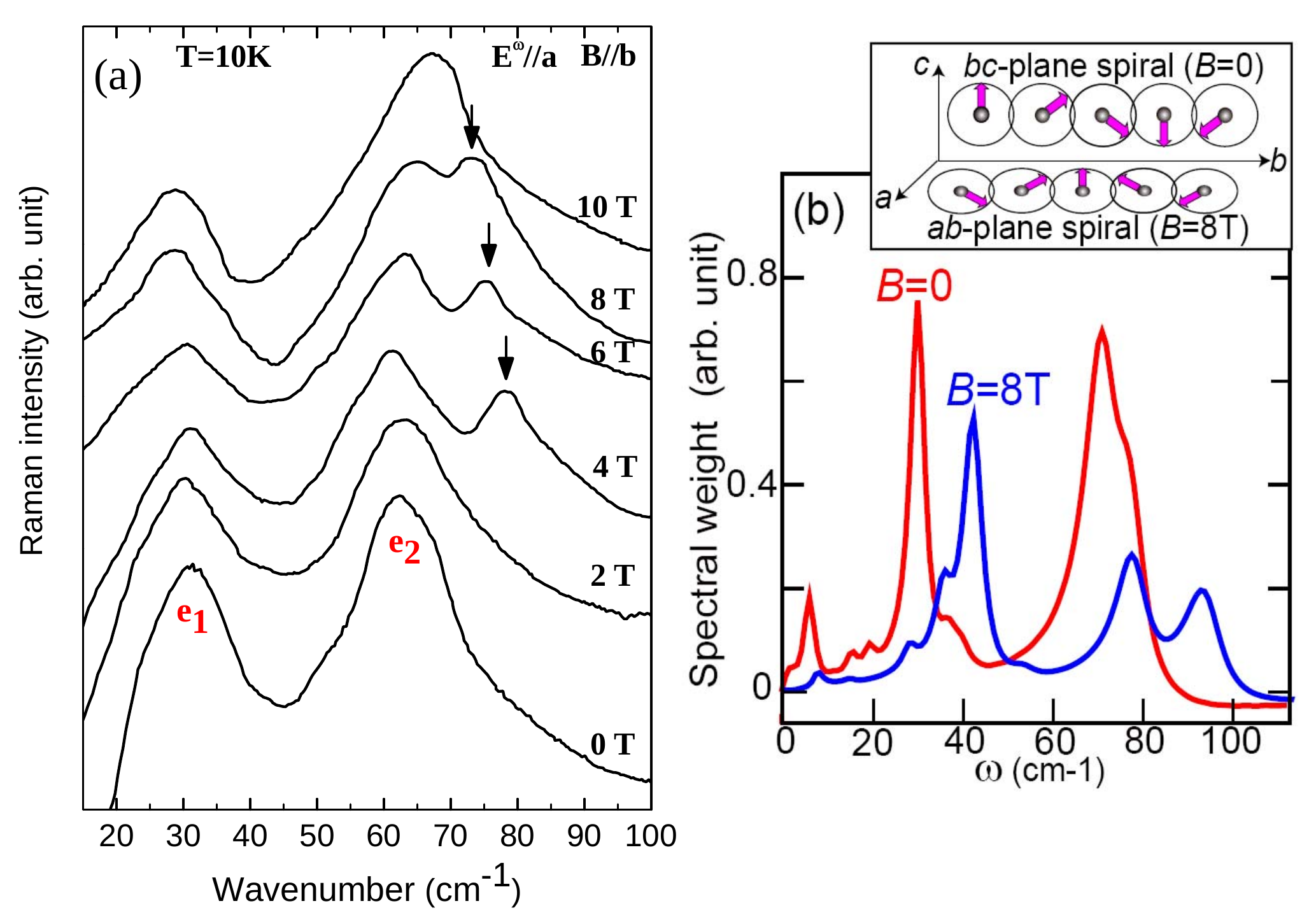}\\
\caption{\label{Figure3} 	 
(a) Magnetic field dependence (\textbf{B}//\textit{b}) of Raman spectra for a polarization of light \textbf{E$^\omega$}//\textit{a}. Two electromagnon modes e$_1$ and e$_2$ are  measured at 30 cm$^{-1}$ and 60 cm$^{-1}$, respectively. An additional  peak appears at 4 T around 78 cm$^{-1}$ when the spiral spin flips from the $bc$ to the $ab$ plane. (b) Sketch of the magnetic spiral structure and calculated spectral weight of electromagnons at B=0 T ($bc$  spiral plane) and B=8 T ($ac$ spiral plane).
}
\end{figure}

We have also performed magnetic measurements along the \textit{b} axis.  
Figure~\ref{Figure3}a shows the Raman spectra measured on the \textit{ac} TbMnO$_3$ sample excited by an electric field of light \textbf{E$^{\omega}$}//a and with a magnetic field \textit{B} along the \textit{b} axis.
Increasing \textit{B} from 0 to 4 T, the Raman spectra remain unchanged. Above 4 T, the frequency of the e$_1$ and e$_2$ electromagnons is weakly shifted. An additional peak appears at 4 T around 78~cm$^{-1}$ when the spin spiral flips from the \textit{bc} to the \textit{ab} plane and shifts down to 68~cm$^{-1}$ at 10 T. Figure~\ref{Figure3}b compares the calculated spectral weight of electromagnons for a magnetic field B=0T (\textit{bc} spiral plane) and B=8 T (\textit{ab} spiral plane). The calculations are performed in the same way as in Ref. \onlinecite{Mochizuki} using a system with 20$\times$20$\times$6 sites. The model parameters for TbMnO$_3$ can be found in Ref. \onlinecite{Pimenov3}. A peak shows up clearly around 90~cm$^{-1}$ for B=8 T when the spin spiral flips in the \textit{ab} plane. It comes from the exchange and easy-plane anisotropy terms. The splitting is caused by the spin-structure modulation induced by the applied magnetic field. Based on the Heisenberg model, this additional  peak is interpreted as an electromagnon resonance. 

In summary, our measurements show the dehybridization of the electromagnons with a magnetic field along \textit{c} and reveal the spectral weight transfer from the electromagnons to the magnon modes. The phonon modes are also strongly impacted. The Mn-O distance appears as the crucial parameter to understand the ferroelectricity and the polar activity of the electromagnons. 
     
\par
The authors would like to thank R. P. S. M. Lobo for helpful discussions.

\end{document}